\documentclass{kluwer}

\newdisplay{guess}{Conjecture}
\usepackage{graphicx}

\newcommand{\Teff}{\ensuremath{T_{\rm eff}}}                            
\newcommand{\logg}{\ensuremath{\log g}}                                 
\newcommand{\kms}{\,km\,s$^{-1}$}                           
\newcommand{\Msun}{\ensuremath{\,{\rm M}_\odot}}            
\newcommand{\Rsun}{\ensuremath{\,{\rm R}_\odot}}            
\newcommand{\degr}{\ensuremath{^{\circ}}}                   

\begin{document}                                                                                   
\begin{article}
\begin{opening}         
\title{Eclipsing Binaries in Open Clusters} 
\author{John \surname{Southworth}\email{jkt@astro.keele.ac.uk}}
\author{Jens Viggo \surname{Clausen}\email{jvc@astro.ku.dk}}  
\runningauthor{Southworth \& Clausen}
\runningtitle{Eclipsing binaries in open clusters}
\institute{Niels Bohr Institute, Copenhagen University, Denmark.}

\begin{abstract}
The study of detached eclipsing binaries in open clusters can provide stringent tests of theoretical stellar 
evolutionary models, which must simultaneously fit the masses, radii, and luminosities of the eclipsing stars 
and the radiative properties of every other star in the cluster. We review recent progress in such studies
and discuss two unusually interesting objects currently under analysis. GV\,Carinae is an A0\,m + A8\,m binary 
in the Southern open cluster NGC\,3532; its eclipse depths have changed by 0.1\,mag between 1990 and 2001, 
suggesting that its orbit is being perturbed by a relatively close third body. DW\,Carinae is a high-mass 
unevolved B1\,V + B1\,V binary in the very young open cluster Collinder\,228, and displays double-peaked 
emission in the centre of the H$\alpha$ line which is characteristic of Be stars. We conclude by pointing out 
that the great promise of eclipsing binaries in open clusters can only be satisfied when both the binaries 
and their parent clusters are well-observed, a situation which is less common than we would like.
\end{abstract}
\end{opening}           


\section{Eclipsing binaries in open clusters}  

Detached eclipsing binary stars (dEBs) are of fundamental importance to stellar physics because they are, 
apart from the few closest objects to the Earth, the only stars for which we can accurately measure basic 
quantities such as mass, radius and surface gravity \cite{and91}. The realism and reliability of the current 
generation of theoretical stellar evolutionary models ranks as one of the great achievements of modern 
astrophysics, but this success would have been much more difficult without the ability to check the effects  
of particular physics against the accurate physical properties of stars in dEBs. 

Given good photometric and spectroscopic data, it is possible to derive masses and radii of stars in a dEB 
to accuracies better than 1\% and surface gravities to within 0.01\,dex \cite{me5}. However, the predictions 
of theoretical models can usually match even properties as accurate as this, both because of their 
sophistication and because there are several important unconstrained parameters, e.g., metal and helium 
abundance and age. More constraints are needed to investigate the success or otherwise of a number of 
physical parameters which are only very simplistically treated in theoretical models, e.g., convective core 
overshooting, mass loss, mixing length and rotational effects. For example, small changes in the mixing 
length can change the derived ages of the oldest globular clusters, which constrain the age of the 
Universe, by 10\% \cite{cha95}.

An answer to this problem is to study dEBs which are members of stellar clusters \cite{me1,tho01}. Because 
the dEB and the other cluster members have the same age and chemical composition, theoretical models must be 
able to simultaneously match the masses, radii and luminosities of the two stars in the dEB and the radiative 
parameters of every other cluster member, for one age and chemical composition. This allows much more detailed 
tests to be made of the success or otherwise of different physical ingredients in models. 
Alternatively, if the cluster is poorly studied, a comparison of the properties of the dEB with model 
predictions allows the cluster metal abundance and age to be derived.

Detached eclipsing binaries are also excellent distance indicators \cite{cla04}, through a variety of methods 
such as using bolometric corrections or surface brightness calibrations (see \opencite{me4} for a detailed 
analysis). dEBs in clusters therefore give an accurate distance to the cluster without the problems and 
theoretical dependence which affect the main sequence fitting method. 

An unique advantage of studying dEBs in clusters is that it can be possible to place four or more stars with 
the same age and chemical composition onto one mass--radius or \Teff--\logg\ plot, if several dEBs are 
members of one cluster (\opencite{me1}; \citeyear{me3}). Some Galactic open clusters are known to contain 
four or more dEBs, e.g., NGC\,7086 (Robb et al., these proceedings).

\section{GV\,Carinae in NGC\,3532}

\begin{figure}
\includegraphics[width=0.49\textwidth]{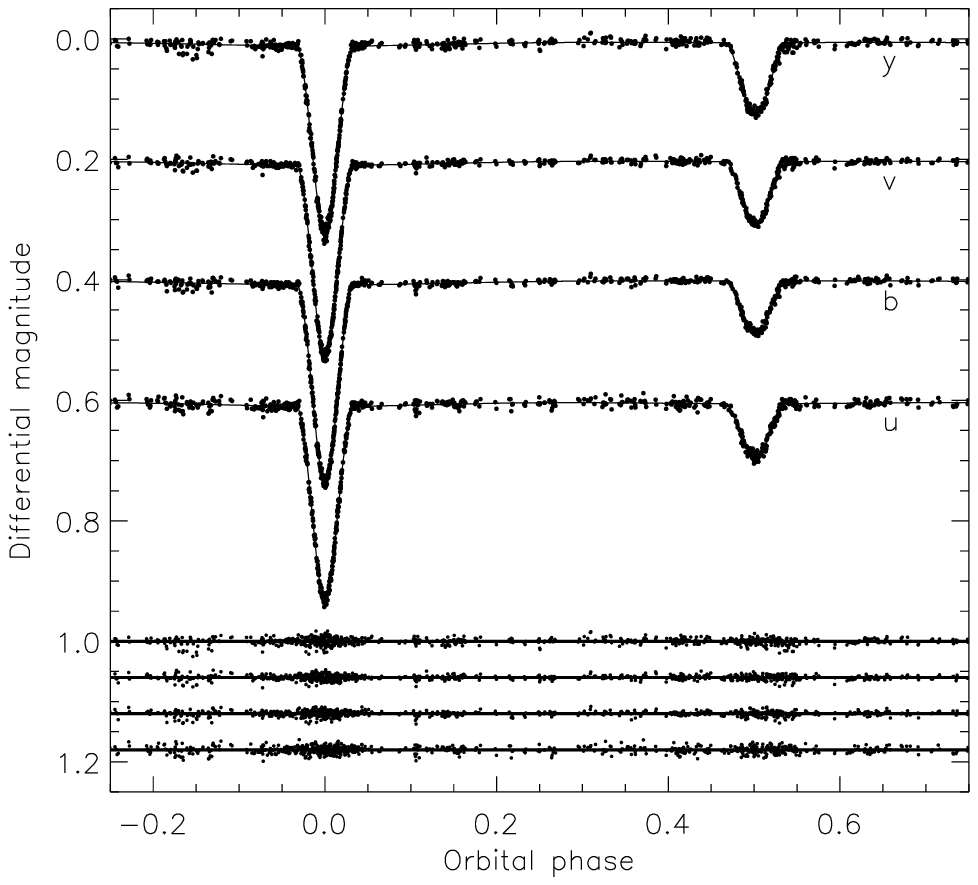}
\includegraphics[width=0.49\textwidth]{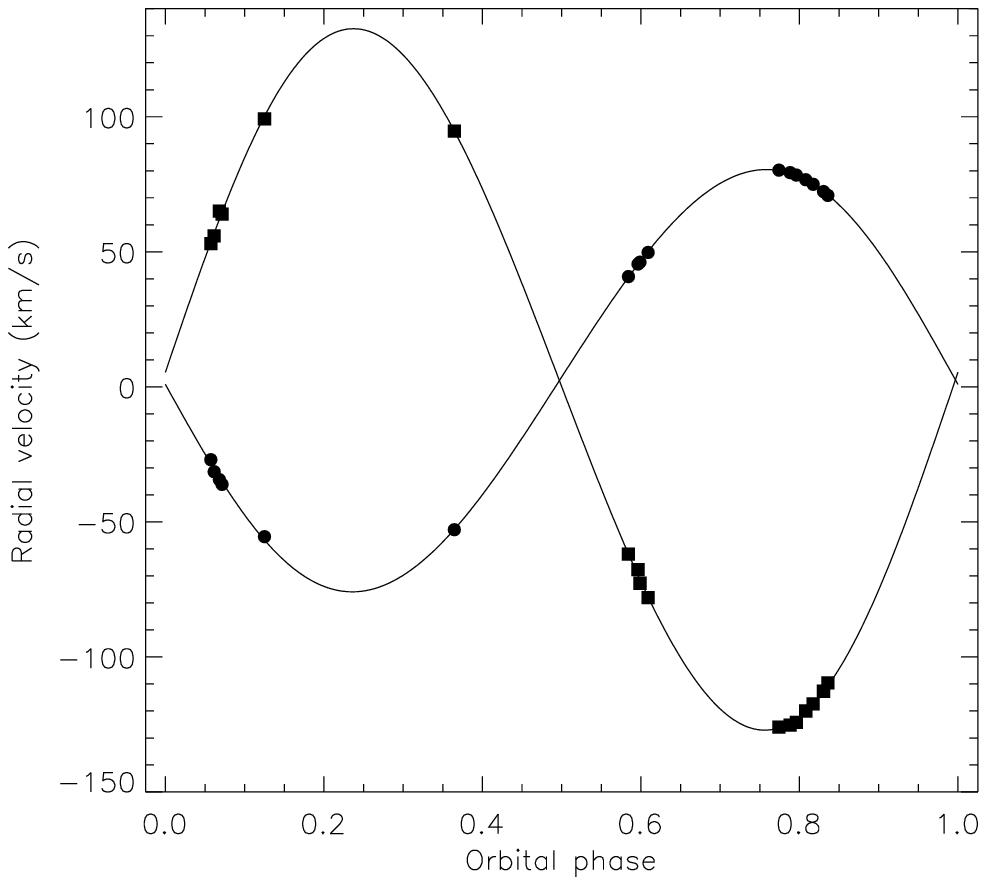}
\caption{The observed light curves (left) and radial velocity curves 
(right) of GV\,Car with the best-fitting models shown using solid lines.}
\label{fig:gv:lcrv}
\end{figure}

GV\,Car ($m_V = 8.9$, $P = 4.29$\,d) is a member of the nearby open cluster NGC\,3532 and contains two 
metallic-lined A\,stars. It displays apsidal motion with a period of $U \approx 300$\,yr.

Complete Str\"omgren $uvby$ light curves, with 775 observations in each passband, were obtained at the 
Str\"omgren Automated Telescope (ESO La Silla) in 1987--1991, with additional data from 2002--2004 
(Fig.~\ref{fig:gv:lcrv}). These light curves have been analysed using the {\sc ebop} code \cite{pop81,etz75} 
and the Monte Carlo error analysis algorithm implemented in {\sc jktebop} (\opencite{me2}; \citeyear{me3}). 

Spectroscopic observations of GV\,Car were obtained in 2001--2004 using the {\sc feros} \'echelle 
spectrograph at the 1.5\,m and 2.2\,m telescopes at ESO La Silla. Radial velocities have been derived by 
cross-correlating spectra of GV\,Car from the 4360--4520\,\AA\ \'echelle order against a spectrum of GV\,Car 
taken at the midpoint of a secondary eclipse. They have been fitted with an eccentric orbit using {\sc sbop} 
(written by P.~B.\ Etzel), which is shown in Fig.~\ref{fig:gv:lcrv}.

\begin{figure}
\includegraphics[width=\textwidth]{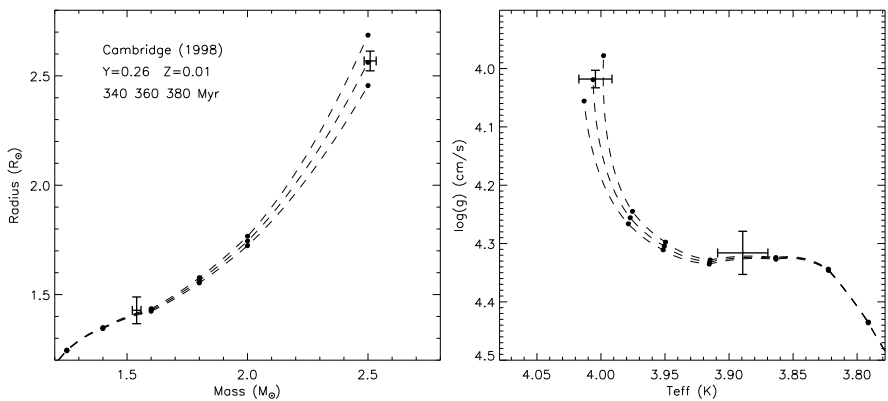}
\caption{Mass--radius and \Teff--\logg\ comparison plots between the properties of the 
components of GV\,Car and the predictions of the Cambridge theoretical models.}
\label{fig:gv:model}
\end{figure}

The spectra have been modelled using the {\sc uclsyn} synthesis code (see \opencite{me1} for references), 
giving ${\Teff}_{\rm A} = 10\,100 \pm 300$\,K and ${\Teff}_{\rm B} = 7750 \pm 350$\,K, consistent with the 
$uvby\beta$ colours of the system and the flux ratios found in the light curve analysis. 

The masses and radii of GV\,Car are $M_{\rm A} = 2.51 \pm 0.03$\Msun, $M_{\rm B} = 1.54 \pm 0.02$\Msun, 
$R_{\rm A} = 2.57 \pm 0.05$\Rsun\ and $R_{\rm B} = 1.43 \pm 0.06$\Rsun. These are well fitted by the Cambridge 
theoretical models \cite{pol98} for an age of $360 \pm 20$\,Myr and a metal abundance of $Z = 0.01$ 
(Fig.~\ref{fig:gv:model}). This age is in good agreement with, and more precise than, main-sequence-fitting 
estimates \cite{gon02}. Our value for the metal abundance is the first published estimate for NGC\,3532.

But we do not yet understand GV\,Car fully; whilst its eclipses were 0.33 and 0.12 mag deep in 1987, they had 
shallowed to 0.22 and 0.08 mag in depth in 2004. This change can be explained by either an increase in 
third light (implying a companion which is itself variable) or a decrease of about 3\degr\ in orbital inclination 
(which suggests a perturbed orbit). The latter explanation seems more likely, but because there is no other 
evidence of a third star in the system it must have a low mass or be a compact object. Further observations 
will be required to fully understand this interesting system.

\section{DW\,Carinae in Collinder\,228}

\begin{figure}
\includegraphics[width=0.49\textwidth]{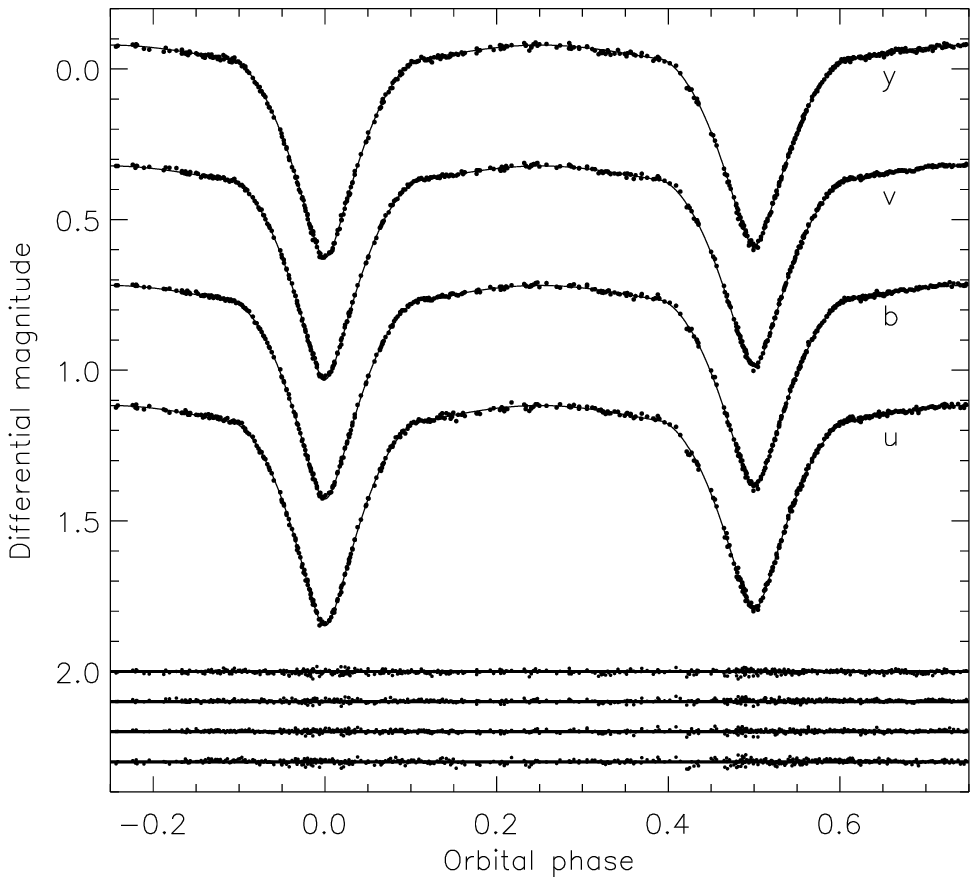}
\includegraphics[width=0.49\textwidth]{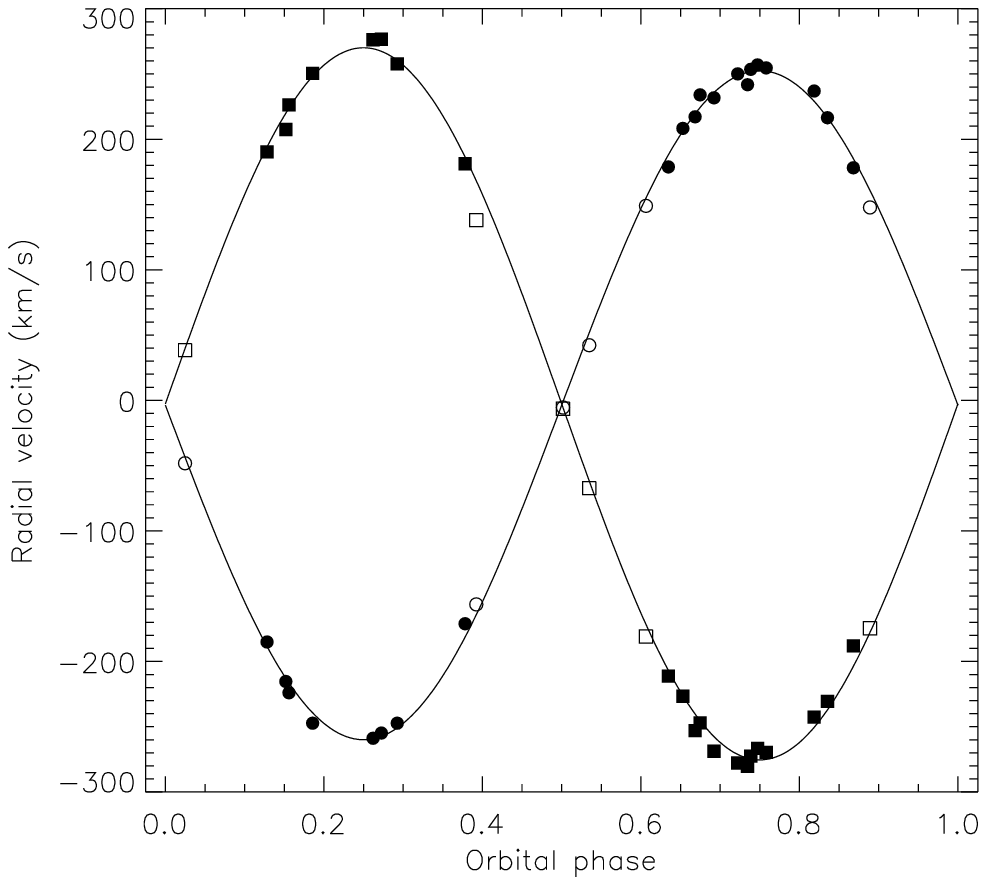}
\caption{Observed light curves (left) and radial velocity curves (right) of DW\,Car with 
best-fitting models shown using solid lines. Radial velocity observations for the primary and secondary stars 
are indicated using circles and squares, respectively, and rejected observations are shown using open symbols.}
\label{fig:dw:lcrv} 
\end{figure}

DW\,Car ($m_V = 9.7$, $P = 1.33$\,d) is a high-mass dEB in the young open cluster Cr\,228. Str\"omgren $uvby$ 
light curves, 518 points in each passband, were obtained as with GV\,Car and modelled using the 2003 
version of the Wilson-Devinney code \cite{wil71} (Fig.~\ref{fig:dw:lcrv}).

The radial velocity analysis of DW\,Car is difficult because the spectra have very few features. Apart from the 
hydrogen lines, which do not give reliable velocities \cite{and75}, there are only four He\,I spectral lines of 
reasonable strength. These lines have been analysed individually using cross-correlation, the {\sc todcor} 
algorithm \cite{zuc94}, Gaussian fitting and spectral disentangling \cite{sim94}. Good results have been obtained 
for disentangling and Gaussian fitting, whilst cross-correlation is significantly affected by line blending. 
Circular orbits were fitted using {\sc sbop}, and the orbit from fitting the He\,I $\lambda$4471 line with a 
double Gaussian is shown in Fig.~\ref{fig:dw:lcrv}.

The resulting masses and radii of DW\,Car are $M_{\rm A} = 11.4 \pm 0.2$\Msun, $M_{\rm B} = 10.7 \pm 0.2$\Msun, 
$R_{\rm A} = 4.52 \pm 0.07$\Rsun\ and $R_{\rm B} = 4.39 \pm 0.07$\Rsun. Str\"omgren index calibrations and the 
flux ratio from the light curves give ${\Teff}_{\rm A} = 27\,500 \pm 1000$\,K and ${\Teff}_{\rm B} = 26\,750 \pm 
1250$\,K. These parameters are acceptably fitted by the predictions of the Cambridge models using the $Z = 0.03$ 
ZAMS, but no conclusions can be drawn from this until definitive values for the radii are obtained.

\begin{figure}
\includegraphics[width=\textwidth]{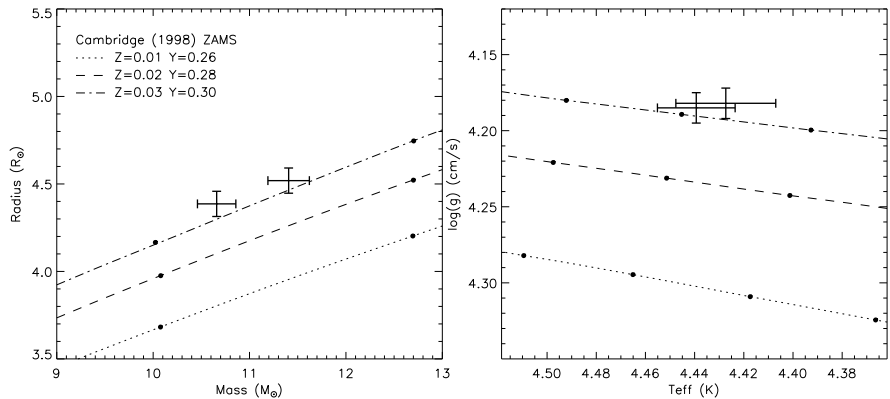}
\caption{Mass--radius and \Teff--\logg\ comparison plots between the properties of the 
components of DW\,Car and the predictions of the Cambridge theoretical models.}
\label{fig:dw:model} 
\end{figure}

DW\,Car shows a double-peaked emission line at H$\alpha$ with a sharp central absorption characteristic of a 
Be star. The line profile does not change with the orbital motion so must come from circumbinary 
rather than circumstellar matter, as 
expected given the closeness of the stars to each other. DW\,Car is a very young system and the rotational 
velocities are `only' about 170\kms. These two facts are very unusual for the Be phenomenon, which is thought 
to increase slightly with age and only be present in stars which are rotating at above 70--80\% of their 
critical velocities \cite{por03}.

\section{Where next?}

The study of dEBs in open clusters has been shown to be an excellent way to determine the parameters of clusters 
by comparison with theoretical stellar models \cite{me2}, but the goal of simultaneously fitting models to both 
the cluster and the two stars in the dEB remains elusive. The main problems are that definitive observations of 
both the cluster and the dEB requires extensive telescope time and that the clusters are often too sparsely 
populated to be useful. 

Clusters containing several dEBs are excellent targets for study, because accurate fundamental parameters can be 
found for four or more stars with the same age, chemical composition and distance, and using the same photometric 
CCD observations. Several Galactic open clusters, in both the Northern and Southern hemispheres, are known to 
contain at least four high-mass dEBs, and full studies of these should provide excellent tests of theoretical 
models.

\theendnotes 

\end{article}
\end{document}